# Drastic changes of electronic structure and crystal chemistry upon oxidation of $Sn^{II}_2TiO_4E2$ into $Sn^{IV}_2TiO_6$: an ab initio study.

Samir F. Matar[a*], Mario Maglione[a], Michel Nakhl[b], Charbel Kfoury[c], Jean Etourneau[a]

[a] CNRS, ICMCB, Université de Bordeaux, Pessac, France

[b] Plateforme de Recherche en Nanosciences et Nanotechnologies, LCPM, Université Libanaise, Fanar, Liban.

[c] Faculté de Génie, Université Libanaise, Fanar, Liban.

*corresponding author: Samir.Matar@icmcb.cnrs.fr

Abstract.

*From DFT based calculations establishing energy-volume equations of state and electron localization mapping, the electronic structure and crystal chemistry changes from $Sn_2TiO_4$ to $Sn_2TiO_6$ by oxidation are rationalized; the key effect being the destabilization of divalent tin $Sn^{II}$ towards tetravalent state $Sn^{IV}$ leading to rutile $Sn_2TiO_6$ as experimentally observed. The subsequent electronic structure change is highlighted in the relative change of the electronic band gap which increases from ~1eV up to 2.2 eV and the 1.5 times increase of the bulk modulus assigned to the change from covalently $Sn^{II}$ based compound to the more ionic $Sn^{IV}$ one. Such trends are also confronted with the relevant properties of black $Sn^{II}O$ characterized by very small band gap.*

## 1. Introduction

Divalent tin $Sn^{II}$ versus isoelectronic $Pb^{II}$ is less encountered in solid state chemistry especially in oxides; ex. perovskite $Pb^{II}TiO_3$ exists whereas $Sn^{II}TiO_3$ has not been prepared as yet in spite of several experimental works [1] and theoretical propositions of its existence and predicted ferroelectric properties [2]. Such instability of $SnTiO_3$ could originate from the difficulty of stabilizing divalent tin in the A site of $ABO_3$ perovskite; contrary to larger $Pb^{II}$. However $Sn^{II}$ in different coordination can be found as in *fordite* mineral $Sn^{II}Nb^{V}_2O_6$ (cf. [3] and therein cited works). It was also stabilized by Kumada et al. [4] in $Sn^{II}_2Ti^{IV}O_4$ ternary which crystallizes in a rare occurrence with the $Pb_3O_4$ ($Pb^{II}_2Pb^{IV}O_4$) tetragonal structure [5]. Both $Sn^{II}$ and $Pb^{II}$ are characterized respectively by $5s^2$ and $6s^2$ electron lone pairs (LP). LP streo-activity was recently reviewed in a large panel of coordinations in fluorides, oxides and oxyfluorides by combining crystal chemistry and theoretical approaches [6,7]. Such stereo-activity is expected to play a role in the properties of $Sn^{II}_2Ti^{IV}O_4$ where the structure keeps relationship with that of tetragonal $TiO_2$ as consisting of endless chains of edge shared *$TiO_6$* octahedra along the c axis, interconnected by $Sn^{II}$. These peculiarities are illustrated in Figure 1 highlighting the chains developing along the tetragonal *c* axis (1a) and their separation by $Sn^{II}$ (1b). The divalent tin atoms are LP-bearing and assembled as tetrahedral motifs developing along the c axis (cf. Fig. 3a). Then a remarkable feature is that the divalent element $Sn^{II}$ ($Pb^{II}$ in $Pb_3O_4$) plays the role of "structural scissor" acting on $TiO_2$ ($PbO_2$) network [7]. This may allow 'tuning' the physical properties for applications. Compounds for photo-catalytic reactions are required for increasing the absorption of visible light. They are usually based on titania modified within the anionic substructure such as by preparing titanium oxy-nitrides [8]. But a modification at the cationic sites may also bring a reduction of the band gap and $Sn_2TiO_4$ was considered for such applications thanks to divalent tin [9].

Above 600°C $Sn^{II}_2Ti^{IV}O_4$ gains weight under oxidation, fixing two oxygen atoms, while changing structure to tetragonal rutile structure type $AO_2$, precisely $Sn_{2/3}Ti_{1/3}O_6$ [4] which is a member of the solid solution $Sn_xTi_{1-x}O_2$ for x = 0.666 investigated by Hirata [10]. From this study it is possible to extract the parameters from this rutile-like cell: a = b = 4.6804Å and c = 3.0955Å, V = 67.81 Å$^3$ from which it is possible to propose an approximate cell for trirutile ordered cell by tripling c parameter: $a_T$ = 4.68 Å, $c_T$ = 9.29 Å, $V_T$ = 203.4 Å$^3$, so that Sn has the property of stabilizing the rutile structure. However the band gap of Sn-modified titania with such 2:1 stoichiometry was not reported and will be discussed comparatively with $Sn^{II}_2Ti^{IV}O_4$ and black SnO herein.

While Ti remains tetravalent throughout the oxidation reaction with persistence of $TiO_2$– like motifs, it can be concluded that tin transforming into tetravalent state, $Sn^{II} \rightarrow Sn^{IV}$ is responsible of major physico-chemical changes involving the substitution of O for E LP and subsequently its stereo activity within the ternary oxide upon transforming $Sn^{II}_2TiO_4$ into $Sn^{IV}_2TiO_6$ rutile type mixed oxide, which perhaps could be ordered via a long annealing around 600°C. This can be expected to affect the properties of electronic structure (band gap) and crystal chemistry (bonding) which are here examined from ab initio with computational methods based on the density functional theory DFT [11,12].

## 2 Computational methodology

Within the accurate quantum mechanical framework of the density functional theory (DFT) we have used the Vienna *ab initio* simulation package (VASP) code [13,14]. VASP allows geometry optimization, total energy calculations and electronic structure description. The projector augmented wave (PAW) method [14,15], is used with atomic potentials built within the generalized gradient approximation (GGA) scheme following Perdew, Burke and Ernzerhof (PBE) [16]. This exchange-correlation XC scheme was preferred to the local density approximation LDA [17] one which is known to be underestimating energy band gaps. The conjugate-gradient algorithm [18] is used in this computational scheme to relax the atoms of the different crystal setups. The tetrahedron method with Blöchl corrections [19] as well as a Methfessel-Paxton [16] scheme was applied for both geometry relaxation and total energy calculations. Brillouin-zone (BZ) integrals were approximated using the special k-point sampling of Monkhorst and Pack [17]. The optimization of the structural parameters was performed until the forces on the atoms were less than 0.02 eV/Å and all stress components less than 0.003 eV/Å$^3$. The calculations are converged at an energy cut-off of 350 eV for the plane-wave basis set with respect to the **k**-point integration with a starting mesh of 6×6×6 up to 12×12×12 for best convergence and relaxation to zero strains. A major outcome of the calculations is the electron localization EL obtained with the ELF function introduced by Becke and Edgecomb [22] as initially devised for Hartree–Fock calculations. Its adaptation to DFT methods was done by Savin et al. [23] as based on the kinetic energy in which the Pauli Exclusion Principle is included: ELF = $(1+ \chi_\sigma^2)^{-1}$ with 0 ≤ ELF ≤1, i.e. it is a normalized function. In this expression the ratio $\chi_\sigma = D_\sigma/D_\sigma^0$, where $D_\sigma = \tau_\sigma -\nabla s - \frac{1}{4}(\nabla\rho_\sigma)^2/\rho_\sigma$ and $D_\sigma^0 = 3/5 (6\pi^2)^{2/3}\rho_\sigma^{5/3}$ correspond respectively to a measure of Pauli repulsion ($D_\sigma$) of the actual system and to the free electron gas repulsion ($D_\sigma^0$) and $\tau_\sigma$ is the kinetic energy density. We mainly consider the 3D iso-surfaces enclosing the electrons of each atomic constituent. Such 3D representations are helpful to visualize post lanthanide ions, here $Sn^{II}$ lone pair development and stereo-activity.

## 3 Geometry optimization and energy volume equations of states

The experimental observation by Kumada et al [4] of $Sn^{II}_2TiO_4$ transformation into rutile type mixed oxide calls for considering a model structure respecting the starting stoichiometry of 2 Sn : 1 Ti. This transformation in rutile like form is ensured by the existence of the solid solution $Sn_xTi_{1-x}O_2$ as indicated above [10] and perhaps as trirutile form. Such ratio is also observed in trirutile aristotype $FeTa_2O_6$ [24] which crystallizes in the same tetragonal $P4_2/mnm$ space group as rutile while allowing for the presence of two different counter anions, i.e. Sn and Ti in 2:1 ratio. We here used the extrapolated $Sn_2TiO_6$ trirutile parameters as starting data for the geometry optimization. Trirutile structure also helped us model the high pressure phase of *fordite* $SnNb_2O_6$ involving a redox reaction with: $Nb^V \rightarrow Nb^{IV}$ and $Sn^{II} \rightarrow Sn^{IV}$ [2]. Fig. 1-c shows the trirutile model structure depicted for $Sn_2TiO_6$ and highlighting the alternating $TiO_6$ and $SnO_6$ edge sharing octahedra in a 3 dimensional arrangement.

Table 1 shows the starting experimental and calculated atomic positions and structure parameters. A fairly good agreement with experiment can be observed for $Sn^{II}_2TiO_4$ and for model $Sn^{IV}_2TiO_6$ with internal parameters within range of those of $Sn_xTi_{1-x}O_2$ [10]. The shortest interatomic distances are compared with experiment for the tetra oxide and they exhibit a good agreement with the trend following $Sn^{II}$ versus $Ti^{IV}$ radii: $r_{SnII}$ = 0.89 Å, $r_{TiIV}$ = 0.60 Å larger for the former, thus leading to d(Sn-O) > d(Ti-O) distance. Note that the difference is much smaller when comparing such distances in trirutile $Sn^{IV}_2TiO_6$ due to the change of Sn valence and then ionic radius ($r_{SnIV}$ = 0.70Å).

The cohesive energies are obtained from the geometry optimized structures shown in Table 1, subtracted from the constituents energies i.e. Sn (-3.746 eV), Ti (-7.753 eV) and $O_2$ (-11.206 eV). In as far as the two compounds have different number of constituents, the comparison is done as per atom: $E_{coh.}(Sn_2TiO_4)$ = -1.82 eV/at., $E_{coh.}(Sn_2TiO_6)$ = -1.65 eV/at.

Sn$_2$TiO$_4$ is identified as more cohesive in its experimentally identified structure and less cohesive in the oxidized phase with rutile based structure. Also we need be aware that our model structure is fully ordered (cf. Fig. 1-c) whereas the true oxide is more likely to a disordered rutile. Nevertheless both compounds have negative enthalpies pointing out to their stability.

The structural modifications are assessed from plotting the energy (E) volume (V), E = $f$(V) calculated around the optimized parameters in Table 1. The curves are shown in Fig. 2. The quadratic behavior is indicative of stable minima. The fits by Birch EOS [25] up to the third order: $E(V) = E_0(V_0) + \frac{9}{8}V_0 B_0[(V_0/V)^{2/3} - 1]^2 + \frac{9}{16}B_0(B'-4)V_0[(V_0/V)^{2/3} - 1]^3$ provides equilibrium parameters: E$_o$, V$_o$, B$_o$ and B$'$ respectively for the energy, the volume, the bulk modulus and its pressure derivative. The obtained values with accurate goodness of fit $\chi^2$ ~10$^{-5}$ / 10$^{-4}$ magnitudes are displayed in the insert of Fig. 2 a,b. The equilibrium volume of Sn$^{II}_2$TiO$_4$ is larger than experiment and the one obtained from energy optimization; this can be assigned to the use of GGA which overestimates volumes contrary to LDA which underestimates them. Also we note that the volume calculated using GGA based PBEsol XC functional is obtained smaller than experiment [9]. However we mainly aim at establishing trends with the changes incurred by the transformation into Sn$^{IV}_2$TiO$_6$.

Trirutile has 2 formula units FU per cell whereas Sn$^{II}_2$TiO$_4$ has 4 FU. Then V$_{FU}$(Sn$^{IV}_2$TiO$_6$) = 101.7 Å$^3$ and V$_{FU}$(Sn$^{II}_2$TiO$_4$) = 106.7 Å$^3$. Such trends point out to volume decrease upon oxidation mainly due to the smaller radius of Sn$^{IV}$ (0.59 Å) versus r(Sn$^{II}$) = 0.89 Å, in spite of the uptake of two additional oxygen atoms. The other large difference between the two compounds is in the magnitude of the respective B$_0$. The bulk modulus defines the resistance of the compound to (isotropic) volume change. With a change from B$_0$(Sn$^{II}_2$TiO$_4$) = 125 GPa to B$_0$(Sn$^{IV}_2$TiO$_6$) = 203 GPa, there is a clear structural effect pertaining to the 3D arrangement of octahedra as edge sharing for both Sn$^{IV}$ and Ti$^{IV}$ in trirutile (Fig. 1 c) whereas chains of *TiO6* edge connecting octahedra are separated by Sn (Fig. 1-a) i.e. the structure is more open.

Lastly for understanding the effect of the presence of Sn$^{II}$O alongside with TiO$_2$ as the Sn$_2$TiO$_4$ structure lets suggest, additional investigation of the EOS and electronic band structure (shown in next section) of SnO itself was undertaken.

SnO exists in two allotropic forms, orthorhombic (red) [26] and tetragonal (black; Romarchite) [27]. Preliminary calculations showed that both have similar electronic band structures; we consider here the latter form for which geometry optimization and subsequent establishment of the EOS were done. Fig. 2c shows the E,V curve and the Birch EOS fit values. Clearly with B$_0$(Sn$^{II}$O) = 99 GPa the oxide is considered as soft, particularly when compared on one hand with SnO$_2$ *cassiterite* with B$_0$~ 220 GPa [28] and with the two ternary oxides on the other hand. These observations confirm the role played by Sn$^{II}$ LP on the mechanical properties either by the presence of Ti$^{IV}$ alongside it in Sn$_2$TiO$_4$ and by the change from Sn$^{II}$ to Sn$^{IV}$ in Sn$_2$TiO$_6$.

### 4. Electron localization function ELF mapping

In the reviews of lone pair LP stereoactivity in different crystal coordinations examined within joint crystal chemistry and DFT approaches [6, 7], 2D and 3D ELF representations allowed following the LP development. Particularly the lone pair is considered as a free electron doublet concentrated in a centroïd, called Ec, which generates around it an electronic cloud, detected in the crystal network as an empty volume, attached to M*, the LP bearing element, here $Sn^{II}$. In this description we label the relevant compounds with adding E, i.e. $Sn_2TiO_4E_2$ and SnOE. We have also shown the paramount importance of the simultaneous presence of SnII and PbII LP's on explaining the properties of the super-ionic conductivity of $PbSnF_4E_2$ [29].

For $Sn_2TiO_4E_2$, SnOE and $Sn_2TiO_6$, Fig. 3 shows the ELF isosurfaces around the atomic constituents with grey 3D volumes and the plane slices –2D views– for the $Sn^{II}$ based compounds ($Sn_2TiO_4E_2$, SnOE) and also $Pb_3O_4E_2$ ($Pb^{II}_2Pb^{IV}O_4$) for the sake of geometric analyses. The ruler is a guide for the eye indicating the color codes: blue, green and red areas corresponding to zero, free electron-like (ELF = ½) and full localization (ELF = 1).

In $Sn_2TiO_4E_2$ (Fig. 3a) the projection onto the basal plane follows that of Fig. 1b. Then the space between the files of $TiO_6$ octahedra is occupied by Sn-LP's oriented toward the empty space forming tetrahedra. The result is the flattening of the LP's which are facing each other. This is opposed to the situation of SnOE (Fig. 3b) whose structure is formed by $SnO_4E$ square pyramids sharing edges and corners, making parallel $[SnOE]_n$ layers packed along [001]. Then Fig. 3b represents a larger E volume development in the free space between the layers versus $Sn_2TiO_4E_2$ where repulsive LP-LP steric influence is observed. This feature is further analyzed in Figs. 3 c showing the ELF 2D slices along similar projections. The numerical exploitations provide $d_{Sn2TiO4E2}$ (Sn-Ec) = 0.82 Å versus $d_{SnO}$(Sn-Ec) =0.97 Å. This further illustrates the difference of LP expansion between the two $Sn^{II}$ based compounds. An important fact must be underlined in the ELF section emphasizing the important difference introduced by the lone pairs in both isostructural compounds $Sn_2TiO_4E_2$ and $Pb_3O_4E_2$. Worthy to note the small electronic isthmus between Sn LP's (Ec-Ec = 2.21 Å, Sn-Sn = 3.720 Å, Sn-Ec = 0.83 Å) in $Sn_2TiO_4E_2$ while there is no contact between Pb LP's in $Pb_3O_4E_2$ (Ec-Ec = 2.95 Å, Pb-Pb = 3.949 Å, Pb-Ec = 0.52 Å).

Another result can be obtained from visual inspection of the localization between Sn and O: whereas for $Sn_2TiO_4E_2$ the ELF's are discontinuous from O to Sn, there is a continuous electron cloud from Sn to O with yellow zones of medium localization in SnO on one hand and also between Sn LP's on the other hand (Fig. 3d). These features of continuous electron flow in SnO allow understanding the vanishingly small band gap (cf. next section) and should be mirrored by differences in the electronic structures as discussed below.

Finally 3D ELF isosurfaces in $Sn_2TiO_6$ (Fig. 3e) undistorted spherical surfaces around LP free $Sn^{IV}$ are observed with the largest ELF volumes around oxygen anions. The difference in electronegativities Ti ($\chi$=1.54) versus Sn ($\chi$= 1.96) explains the absence of ELF around Ti while a small remaining ELF surrounds Sn.

### 5. Electron band structure and site projected density of states (PDOS)

Figs. 4 show the band dispersions along the major directions of the tetragonal Brillouin zone. In all panels the energy *y*-axis has its reference at $E_V$, top of the valence band VB, in so far that all compounds are insulating of small gap semi-conductors. The latter is actually the case of SnO (fig. 4a) which very close to a weak metal with a vanishingly small gap between $\Gamma_{Valence}$ and $M_{Conduction}$ point at the bottom of the conduction band (CB). This results from the largely dispersive bands thanks to the s,p character. Also low energy lying O 2s states are not shown as they were not considered in the valence basis set in the calculations.

In Sn$_2$TiO$_4$ (fig. 4b) the band structure is different due to the presence of Ti which brings localized d states especially at the CB bottom due to the centering of the low filled Ti 3d states within CB (cf. PDOS at Figs. 5). Consequently a band gap opening at $\Gamma$ point occurs with $\Gamma_{Valence}$ -$\Gamma_{Conduction}$ ~1 eV magnitude. The VB shows two parts extending over 6 eV from VB top down to -6 eV corresponding the mixing of the valence states of constituents and a smaller energy windows over 2 eV with mainly Sn s states (cf. Figs. 5).

Comparatively the and gap in Sn$_2$TiO$_6$ (fig. 4c) is almost twice larger than in Sn$_2$TiO$_4$ with $\Gamma_{Valence}$ -$\Gamma_{Conduction}$ ~2 eV due to the tetravalent character of both Sn and Ti in rutile AO$_2$ (A=Sn, Ti). Also the VB shape is changed with separation between the low energy lying s block from -6 to -8 eV and the ~5 eV broad and narrower toward the top.

Then upon going through the series of the three Sn based compounds, there is a progressive change of the electronic band structure thanks to the chemical modifications describes above.

The site projected density of states (PDOS) presented in Figs. 5 mirror the band structure observations. The energy reference is with respected to E$_V$ at the top of the valence band.

    A    SnO (Fig. 5a): E$_V$ is seen here at a shallow DOS minimum corresponding to the vanishingly small band gap between $\Gamma_{Valence}$ and M$_{Conduction}$ (Fig. 4a). The valence band shows correspondingly a continuous DOS shape with prevailing Sn-O PDOS at the VB top. The bottom of CB shows dominant Sn states.

    B-    Sn$_2$TiO$_4$ (Fig. 5b): Similarly to SnO the top of the VB is dominated by low intensity and broad, i.e. dispersive Sn states. This also stands for the s-block from -9 to -6 eV. Then it can be observed that Ti with low intensity PDOS comes as 'a perturbation" of SnO-based compound. On the contrary the CB comprises dominant Ti empty states so that the flat bands at the CB bottom are due to empty Ti d states.

    C-    Sn$_2$TiO$_6$ (Fig. 5c): The VB is dominated by oxygen p-states with large oxygen PDOS intensities throughout VB and up to E$_V$. Here too the CB bottom shows mainly empty Ti d PDOS with a direct gap, $\Gamma_V$- $\Gamma_C$ = ~4 eV. Testing local density approximation LDA [14] potentials led to similar trends but with smaller gap magnitudes due to the overbinding character of the LDA.

## 6. Conclusions

In this work it has been shown that the post lanthanide divalent elements as Sn$^{II}$ and Pb$^{II}$ play the role of "structural scissor" acting on the dioxide network (TiO$_2$, PbO$_2$) network. Significant modifications of the electronic structure are subsequently observed thanks to accurate analyses within the DFT quantum theory of the electron localization function ELF and the electronic band structures. The ELF analysis combined with crystal chemistry of Sn$^{II}$ lone pair LP stereoactivity has particularly shown that LP-LP overlap in SnO provides interpretation of its very small band gap. The Sn$^{II}$ LP stereoactivity changes upon adjoining TiO$_2$ to two Sn$^{II}$O formulae units. Resulting Sn$^{II}_2$TiO$_4$ undergoes a ~1eV band gap opening thus reducing the large band gap of titanate, thence allowing applications. Oxidizing Sn$_2$TiO$_4$ into rutile mixed oxide as evidenced experimentally was modeled here in trirutile structure:

$Sn^{IV}_2TiO_6$ leading to its characterization with a twice larger band gap. Such subtle tuning of band electronic structure thanks to the electronic nature of the p-element (post lanthanide) should trigger further basic research on relevant compounds with LP-bearing ions.

Lastly it is interesting to propose a parallel with the Ruddlesden Popper type phases: $CaO(CaMnO_3)_n$ with n = 1, 2, 3, … where calcium oxide interlayers perovskite like motifs as with the series $Ca_3MnO_4$ (n = 2), $Ca_3Mn_2O_7$ (n = 3), … studied mainly for their magnetic structures (cf. [30] and therein cited works). Here with SnO and within "anti-Ruddlesden-Popper" manner we get: $(SnO)_nTiO_2$. With n = 1 we get $(SnO)TiO_2$ or $SnTiO_3$ proposed formerly from theory in perovskite structure [1] but not yet prepared*. Occurrence of presently studied $(SnO)_2TiO_2$ or $Sn_2TiO_4$ [3] experimentally evidenced occurs with n = 2. Then it would be relevant to examine and model crystal and electronic structures of n = 3: $(SnO)_3TiO_2$ or $Sn_3TiO_5$ which could be brownmillerite-based structure… Such studies are underway.

**Acknowledgements**

We acknowledge intensive and fruitful exchange on electron lone pair with Dr Jean Galy (Emer. CNRS Research Director at LCTS-CNRS-University of Bordeaux). We thank the CNRS-France, the INC-CNRS the *Conseil Régional d'Aquitaine* and MCIA-University of Bordeaux for support. One of us SFM acknowledges facilities provided by the *Plateforme Recherche en Nanoscience & Nanotechnologies* of the Lebanese University (Fanar; Lebanon) where part of the work was done.

**Table 1**

Sn$_2$TiO$_4$ [4]
P4$_2$/mbc # 135
Ti (4*d*) 0 ½ ¼
a = 8.49 (8.54) Å
c = 5.923 (5.97) Å
V / 4 FU = 426.93 Å$^3$ (432.3 Å$^3$)
Total energy: -197.84 eV

___________________________________________________________

| Atom | | x | y | z |
|---|---|---|---|---|
| Sn | 8*h* | 0.1451 (0.146) | 0.1630 (0.166) | 0 |
| O1 | 8*g* | 0.664 (0.664) | 0.164 (0.165) | ¼ |
| O2 | 8*h* | 0.098 (0.097) | 0.621 (0.622) | 0 |

d(Sn-O1) = 2.09 (2.12) Å   d(Sn-O2) = 2.21 (2.22) Å
d(Ti-O1) = 1.97 (1.98) Å  d(Ti-O2) = 1.99 (2.0) Å
___________________________________________________________

Sn$_2$TiO$_6$ this work
Model Trirutile *P4$_2$/mnm* #136
Fe (Ti) (2*a*) 000;
a = 4.68 (4.612) Å
c = 9.29 (9.147) Å
V = 203.45 Å$^3$ (194 Å$^3$)
Total energy: -127.39 eV

___________________________________________________________

| Atom | | x | y | z |
|---|---|---|---|---|
| ~~Ta~~ (Sn | 4*e* | 0 | 0 | 0.333 (0.334) |
| O1 | 4*f* | 0.307 (0.311) | 0.307 (0.311) | 0 |
| O2 | 8*j* | 0.297 (0.299) | 0.297 (0.299) | 0.322 (0.326) |

d(Sn-O1) = 1.93   d(Sn-O2) 1.92
d(Ti-O1) = 1.91  d(Ti-O2) = 1.93

**Figures**

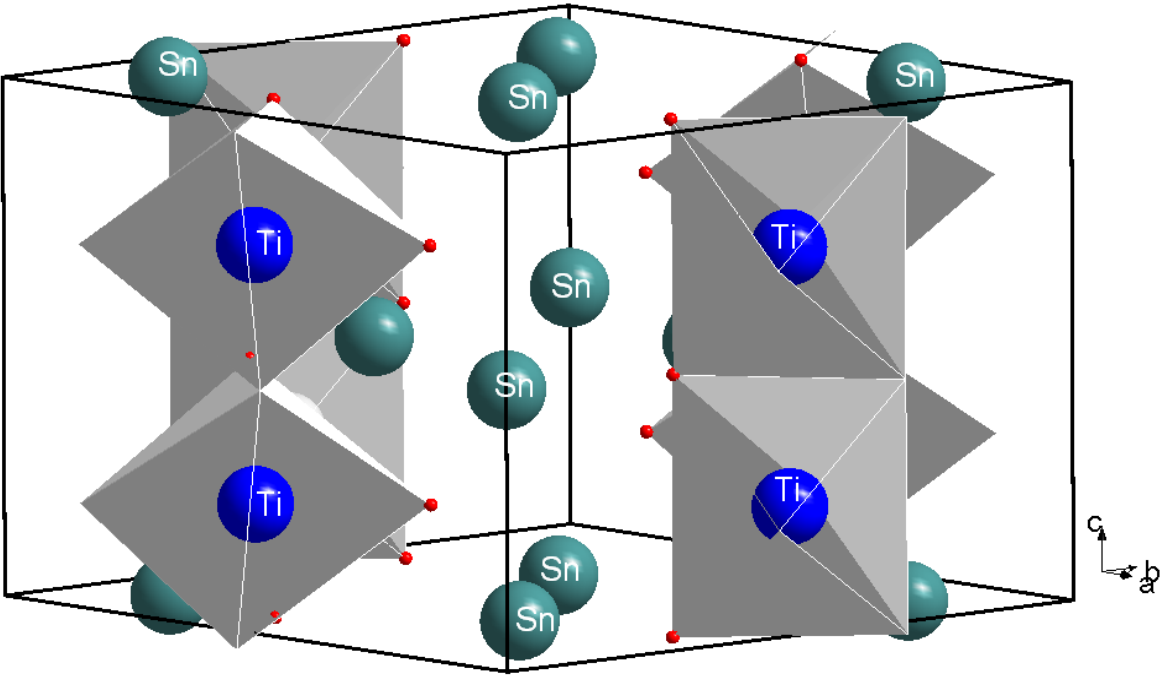

a)

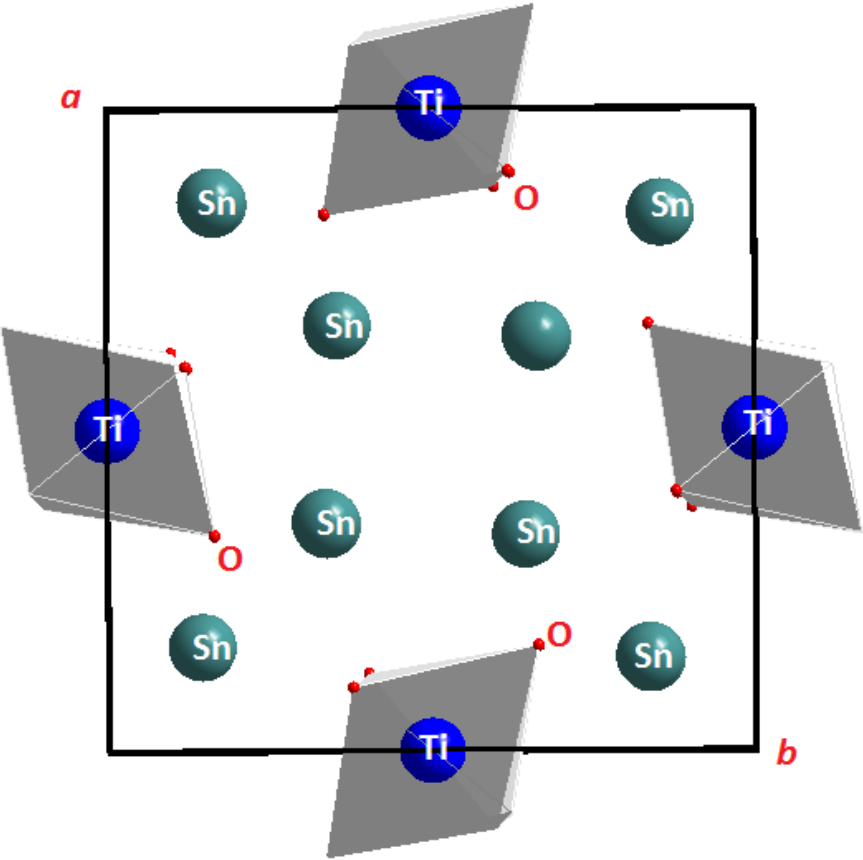

b)

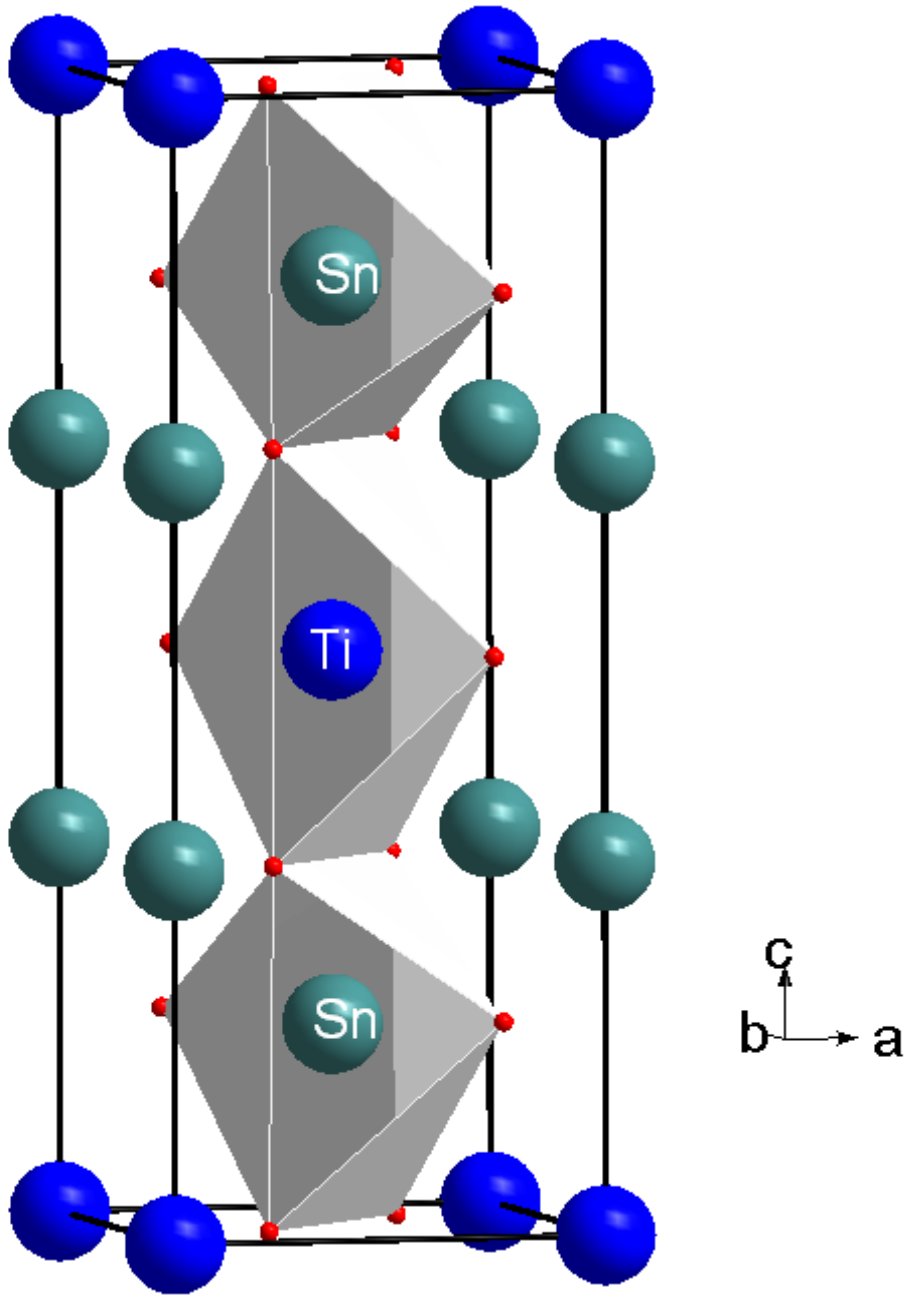

Fig. 1. a and b) Sn$_2$TiO$_4$ structure highlighting the connections of the edge sharing *TiO6* octahedra arranged in infinite chains developing along the *c* axis. c) Trirutile structure of Sn$_2$TiO$_6$ with *TiO6 – SnO6* edge sharing octahedral.

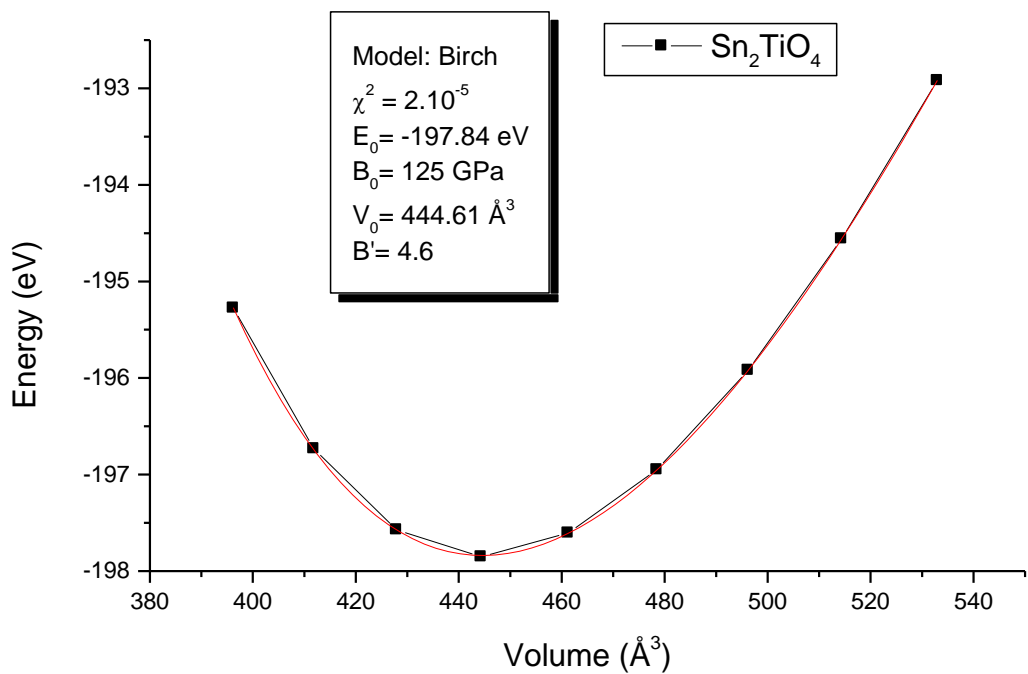

a)

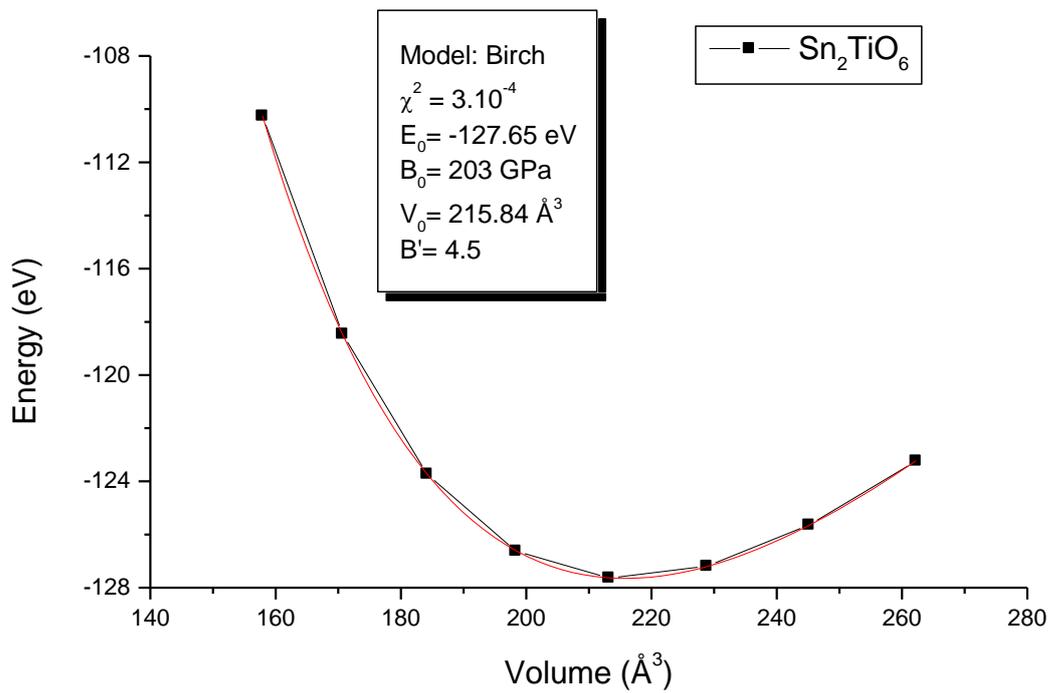

b)

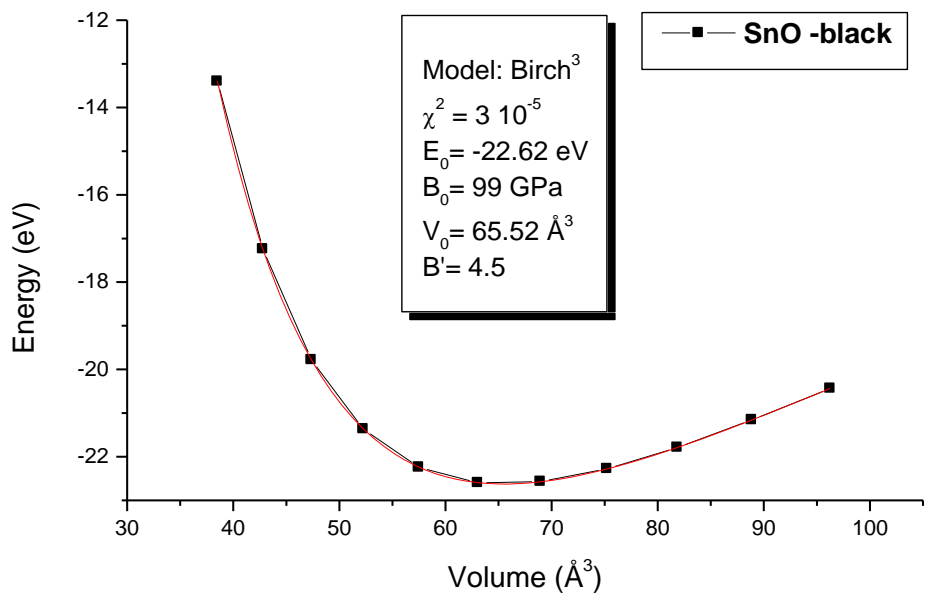

c)

Fig. 2 Energy volume curves of $Sn_2TiO_4$ (a), $Sn_2TiO_6$ (b) and SnO Romarchite (c).

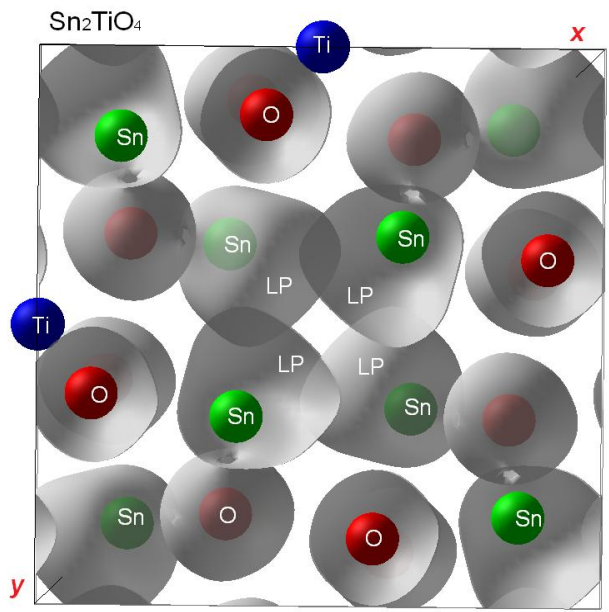
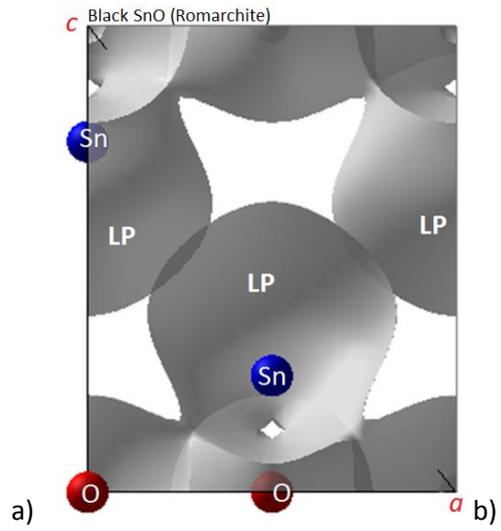

a)                                                                                       b)

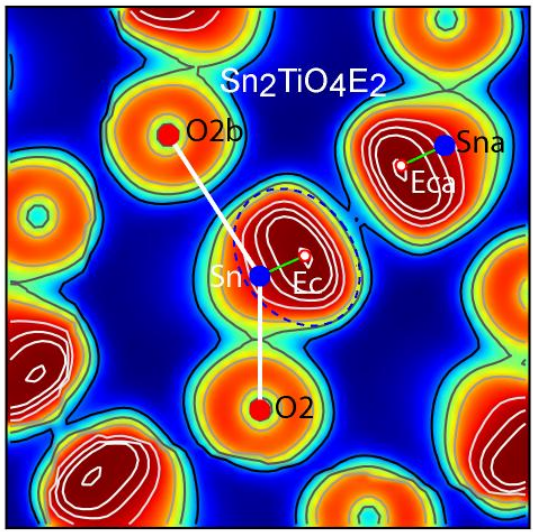
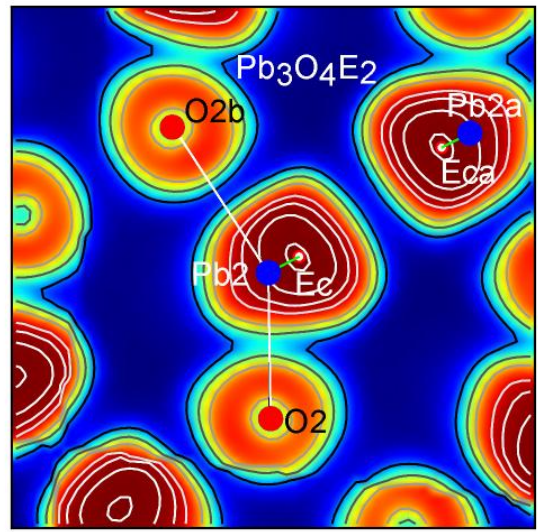

c)

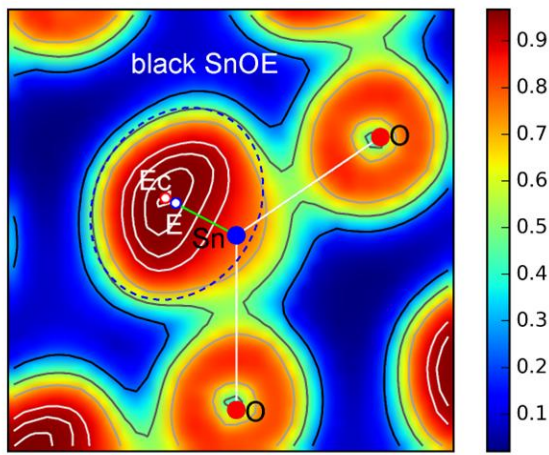
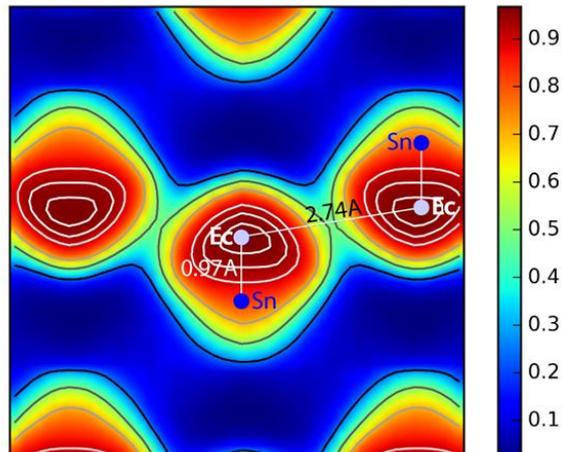

d)

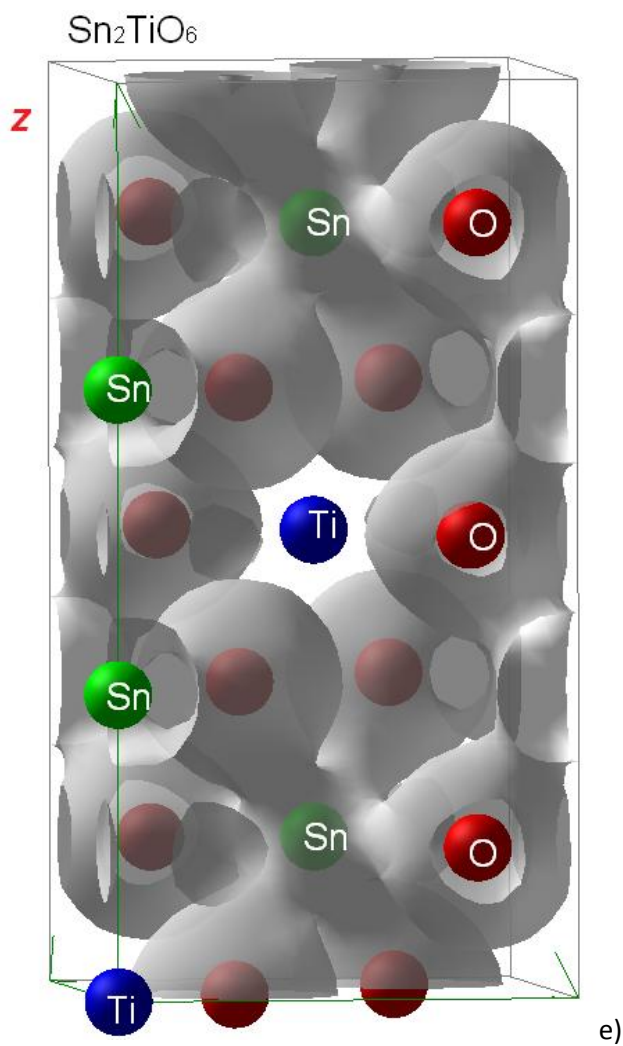

Fig. 3. 3D electron localization function ELF of $Sn_2TiO_4$, SnO and $Sn_2TiO_6$ (a, b, d) highlighting LP $Sn^{II}$ volume development in $Sn_2TiO_4$ and SnO and its absence in $Sn_2TiO_6$. Following the projection in Fig. 1b, the LP bearing $Sn^{II}$ are arranged in tetrahedra along *c*. Sn LP's are oriented toward the empty space between the tetrahedra thus acting as chemical scissors in $TiO_2$.

2D ELF slices with scales between 0 and 1 for zero localization (blue zones) and full localization (red zones) for $Sn_2TiO_4$ (c), and SnO Romarchite (d) (cf. text).

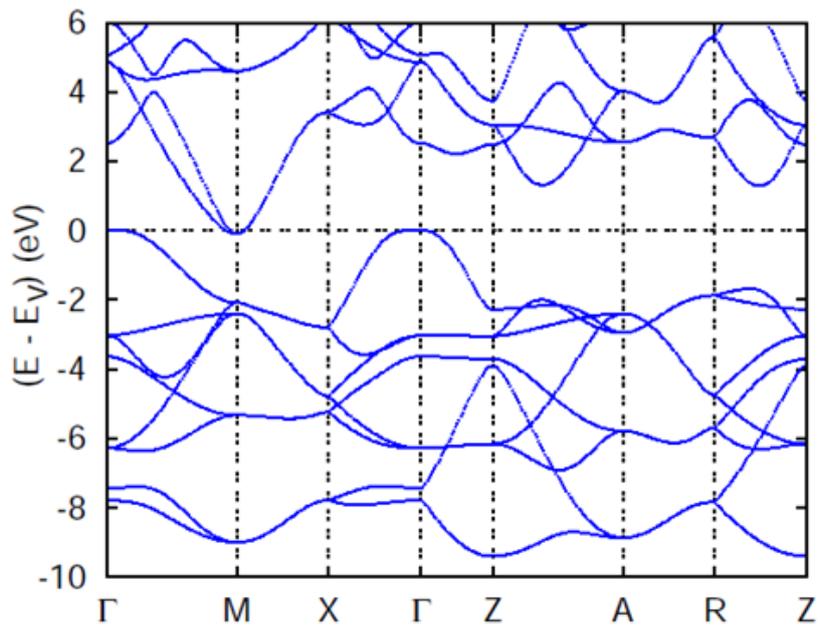

a)

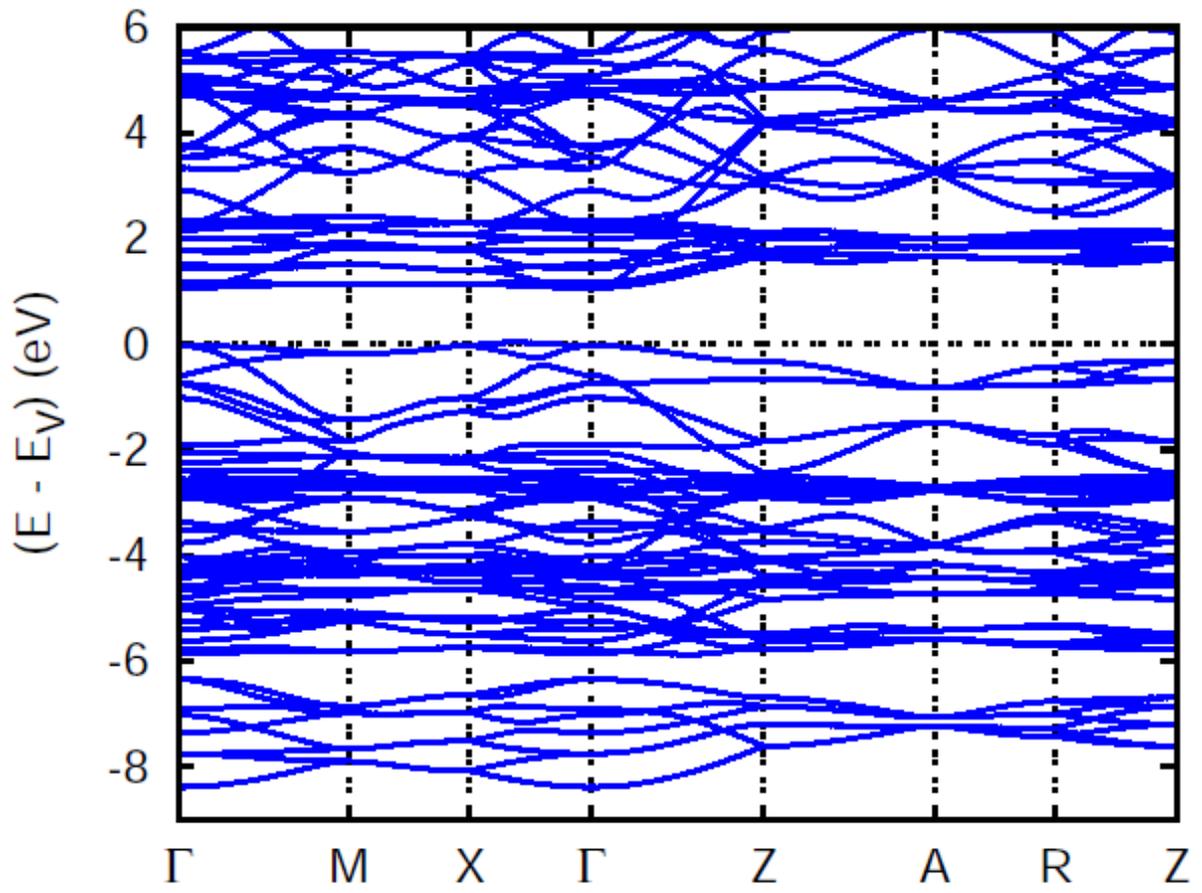

b)

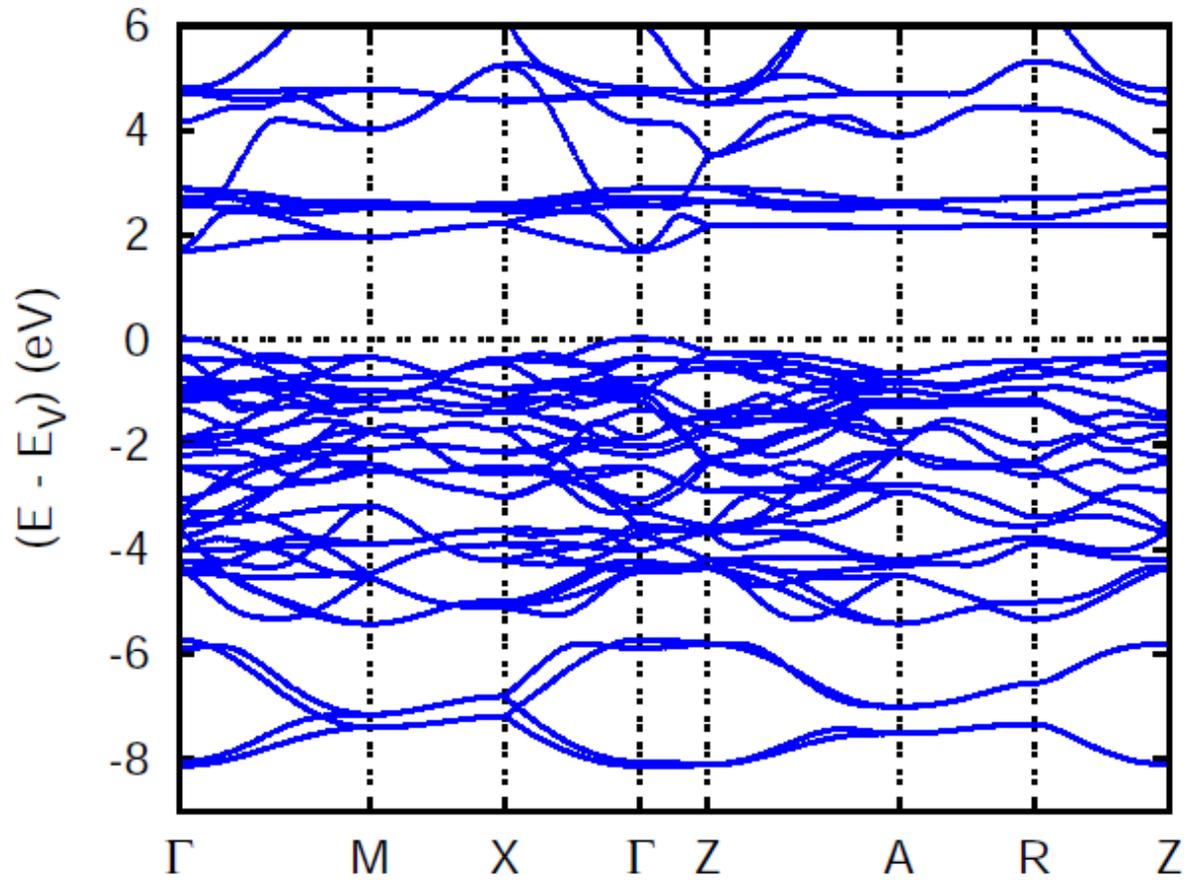

c)

Fig. 4 Electronic band structures of SnO Romarchite (a), $Sn_2TiO_4$ (b) and $Sn_2TiO_6$ (c) along major directions of simple tetragonal Brillouin zone BZ. $\Gamma$ corresponds to the BZ center.

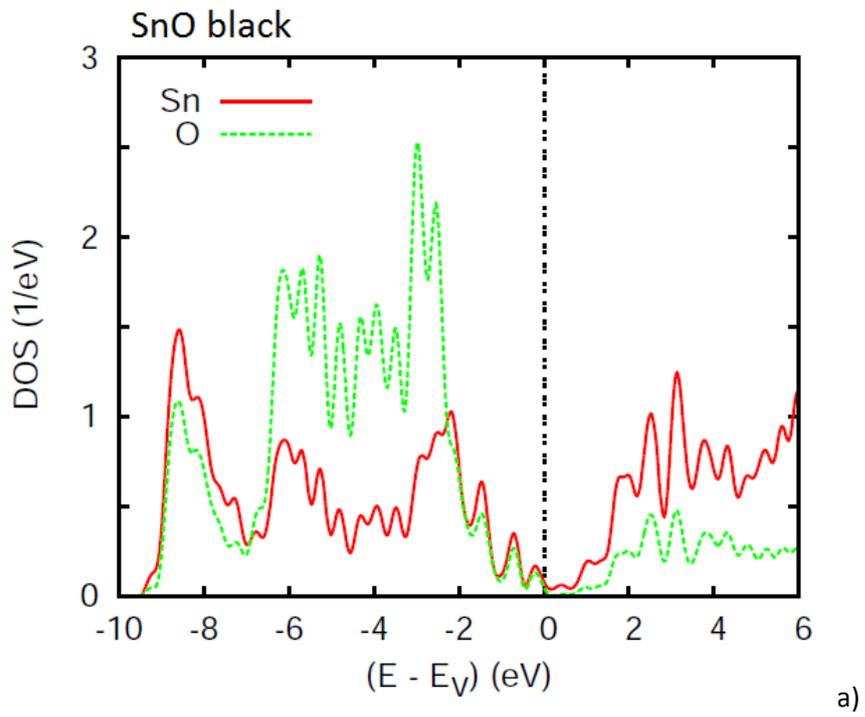

a)

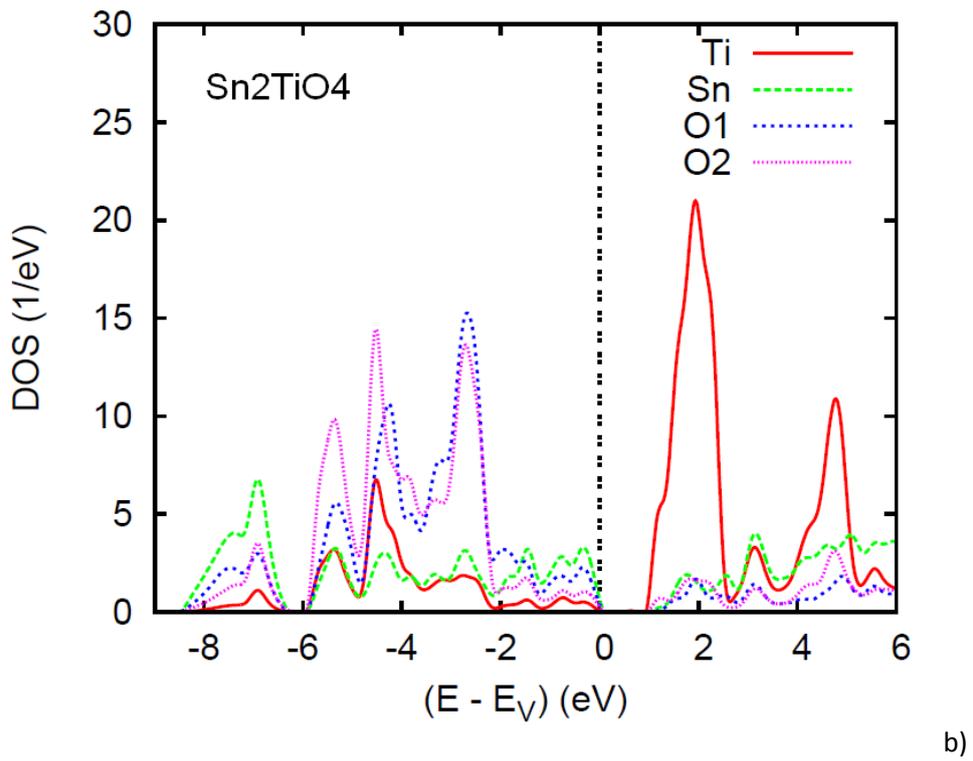

b)

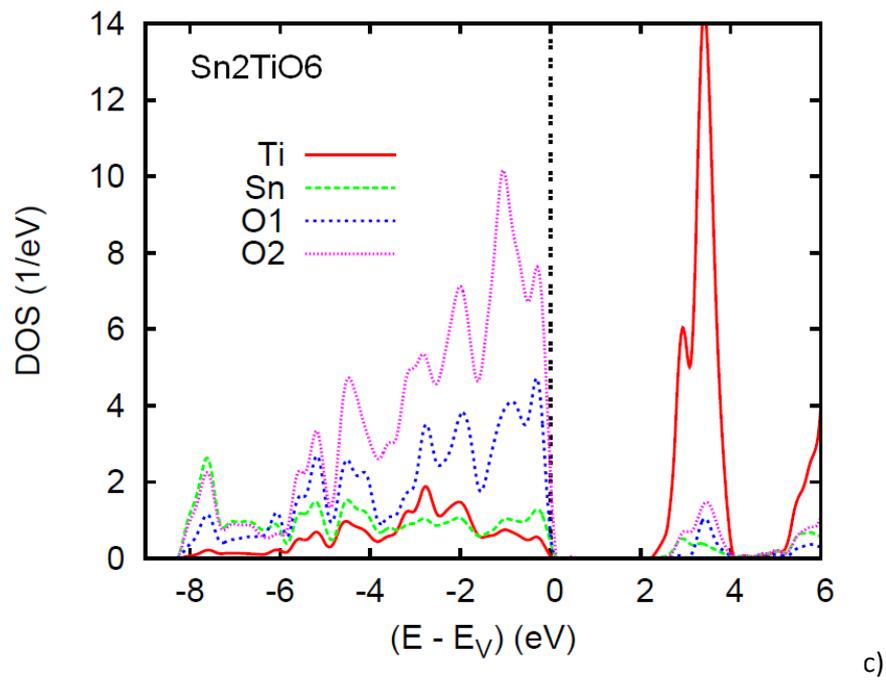

Fig. 5 Site projected density of states of SnO Romarchite (a), $Sn_2TiO_4$ (b) and $Sn_2TiO_6$ (c).**